\documentclass[prc,showpacs,showkeys,superscriptaddress,nofootinbib,floatfix,twocolumn]{revtex4}
\usepackage{color}
\usepackage{amsfonts}
\usepackage{amsbsy}
\usepackage{mathrsfs}
\usepackage{graphicx}
\def\lsim{\mathrel{\rlap{
\lower4pt\hbox{\hskip-3pt$\sim$}}
    \raise1pt\hbox{$<$}}}     
\def\gsim{\mathrel{\rlap{
\lower4pt\hbox{\hskip-3pt$\sim$}}
    \raise1pt\hbox{$>$}}}     
\def\scr#1{\mbox{\scriptsize #1}}
\usepackage{graphicx}
\begin{document}
\title{Transverse-Mass Spectra in Heavy-Ion Collisions at energies
$E_{\scr{lab}} =$ 2--160 GeV/nucleon }%
\author{Yu.B.~Ivanov}\thanks{e-mail: Y.Ivanov@gsi.de}
\affiliation{Gesellschaft
 f\"ur Schwerionenforschung mbH, Planckstr. 1,
D-64291 Darmstadt, Germany}
\affiliation{Kurchatov Institute, Kurchatov
sq. 1, Moscow 123182, Russia}
\author{V.N.~Russkikh}\thanks{e-mail: russ@ru.net}
\affiliation{Gesellschaft
 f\"ur Schwerionenforschung mbH, Planckstr. 1,
D-64291 Darmstadt, Germany}
\affiliation{Kurchatov Institute, Kurchatov
sq. 1, Moscow 123182, Russia}
\begin{abstract}
Transverse-mass spectra of protons, pions and kaons produced in
collisions of heavy nuclei are analyzed within the model of 3-fluid
dynamics. It was demonstrated that this model consistently 
reproduces these spectra in wide ranges of incident energies,  
4$A$ GeV $\lsim E_{\scr{lab}} \lsim$ 160$A$ GeV, rapidity bins and
centralities  of the collisions. In particular, the model describes
the "step-like" dependence of kaon inverse slopes on the incident energy.
The key point of this explanation is interplay of hydrodynamic expansion
of the system with its dynamical freeze-out. 
\pacs{24.10.Nz, 25.75.-q}
\keywords{relativistic heavy-ion collisions, hydrodynamics,
  transverse-mass spectra }
\end{abstract}
%
\maketitle

\section{Introduction}

One of the major goals of high-energy heavy-ion research 
is to explore properties of strongly interacting matter,
particularly its phase structure \cite{goals}.
A great body of experimental data has  been already accumulated by now. 
Theoretical analysis of these data is still in progress. 
Explanation of 
transverse-mass spectra of hadrons (especially kaons) in collisions of
heavy nuclei 
turned out to be one of the most difficult tasks
\cite{Bratkovskaya,Wagner,Hama04}.

Experimental data on transverse-mass spectra of kaons produced in
central Au+Au \cite{E866} and Pb+Pb \cite{mg:04,NA49} collisions reveal
peculiar dependence on the incident energy. The inverse-slope
parameter (so called effective temperature $T$) of these spectra at
midrapidity increases with incident energy in the energy domain of
BNL Alternating Gradient Synchrotron (AGS) and then saturates at
energies of CERN Super Proton Synchrotron (SPS). In Refs. 
\cite{Gorenstein03,Mohanty03} it was assumed that this saturation is
associated with the deconfinement phase transition. This assumption
was indirectly confirmed by the fact that microscopic transport
models, based on hadronic degrees of freedom, failed to reproduce
the observed behavior of the kaon inverse slope \cite{Bratkovskaya,Wagner}.
Hydrodynamic simulations of Ref. \cite{Hama04} succeeded to describe
this behavior. However, in order to reproduce it these hydrodynamic
simulations required incident-energy dependence of the freeze-out
temperature which almost repeated the shape of the corresponding
kaon effective temperature. This happened even in spite of using
an equation of state (EoS) involving a phase transition into the 
quark-gluon plasma (QGP). This way, the puzzle of kaon effective
temperatures was just translated into a puzzle of freeze-out
temperatures. Moreover, results of Ref. \cite{Hama04} imply that
peculiar  incident-energy dependence of the kaon effective
temperature may be associated with dynamics of freeze-out.

In Ref. \cite{3FDpt} it was shown that dynamical description of
freeze-out \cite{3FDfrz}, accepted in the  model of 3-fluid dynamics
(3FD) \cite{3FD,3FDm,3FDflow,3FD-GSI07}, naturally explains the incident energy 
behavior of inverse-slope parameters of transverse-mass spectra observed in
experiment. This freeze-out dynamics, effectively resulting in a
pattern similar to that of the dynamic liquid--gas transition, differs
from conventionally used  freeze-out schemes.
In the brief letter \cite{3FDpt} only 
midrapidity inverse-slope parameters in central collisions were
presented. In the present paper we would like to extend analysis of
Ref. \cite{3FDpt} by presenting transverse-mass spectra themselves
at various impact parameters and rapidities and compare them with
available experimental data in the range from AGS to SPS incident
energies. These spectra are computed with the same set of model parameters 
as that summarized in Ref. \cite{3FD}. In particular, the hadronic EoS 
\cite{gasEOS} with incompressibility $K=210$ MeV is used.

\section{The 3FD Model}
\label{The 3FD Model}

The 3FD model is designed for
simulating heavy-ion collisions in the range from AGS to SPS energies.
Unlike the conventional hydrodynamics, where local instantaneous
stopping of projectile and target matter is assumed, a specific
feature of the dynamic 3-fluid description is a finite stopping
power resulting in a counter-streaming regime of leading
baryon-rich matter.  This counter-streaming is described in terms of 
two interacting baryon-rich fluids,
initially associated with constituent nucleons of the projectile
(p) and target (t) nuclei. In addition, newly produced particles,
populating the midrapidity region, are associated with a baryon-free 
``fireball'' (f) fluid.  

We have started our simulations \cite{3FDpt,3FD,3FDflow}
with a simple hadronic EoS \cite{gasEOS}.
The 3FD model turned out to be able to reasonably
reproduce a large body of experimental data \cite{3FDpt,3FD,3FDflow} in a wide
energy range from AGS to SPS. This was done with a unique set of
model parameters summarized in Ref. \cite{3FD}.
Problems were met in description of the transverse flow \cite{3FDflow}.
The directed flow required a softer EoS at top AGS and SPS energies
(in particular, this desired softening may signal occurrence of the 
phase transition into the QGP).

The transverse-mass spectra are most sensitive to the freeze-out parameters 
of the model. 
In the 3FD model the same freeze-out procedure is applied to 
both the thermal and chemical processes. 
Normalization of the meson spectra is directly related to the freeze-out 
temperature. 
Inverse slopes of the transverse-mass spectra
represent a combined effect of the temperature and collective
transverse flow of the hydrodynamical expansion.
Had it been only the effect of thermal excitation, inverse slopes
for different
hadronic species would approximately equal. 
The collective transverse flow makes them different.
These two effects partially
compensate each other: the later freeze-out occurs, the lower
temperature and the stronger collective flow are.
Nevertheless, inverse slopes turn out to be sensitive
to the instant of the freeze-out.
Below we demonstrate that not only slopes but also normalizations 
of the pion transverse-mass spectra are well described by this 
unified chemical-thermal freeze-out. Situation with kaons is 
more delicate, since we have to take into account the associative 
production of strangeness giving rise to a strangeness 
suppression factor.

Freeze-out procedure adopted in the 3FD model 
was analyzed  in detail in Ref. \cite{3FDfrz}. 
This method of freeze-out can be called dynamical, since the
freeze-out process here is integrated into fluid dynamics through
hydrodynamic equations. The freeze-out
front is not defined just ``geometrically'' on the condition of the
freeze-out criterion met\footnote{The freeze-out criterion 
demands that the energy density of the matter is lower than some value
$\varepsilon_{\scr{frz}}$.} 
but rather is a subject the fluid evolution.
It competes with the fluid flow and not always reaches the place where
the freeze-out criterion is met.
This kind of freeze-out is similar to the model of ``continuous
emission'' proposed in Ref. \cite{Sinyukov02}. There the particle emission
occurs from a surface layer of the mean-free-path width. In our case the
physical pattern is similar, only the mean free path is shrunk to zero.
We would like to mention that recently the continuous emission model
was further microscopically developed  \cite{Sinyukov08,Knoll08}.


%
\begin{figure}[thb]
\includegraphics[width=6.9cm]{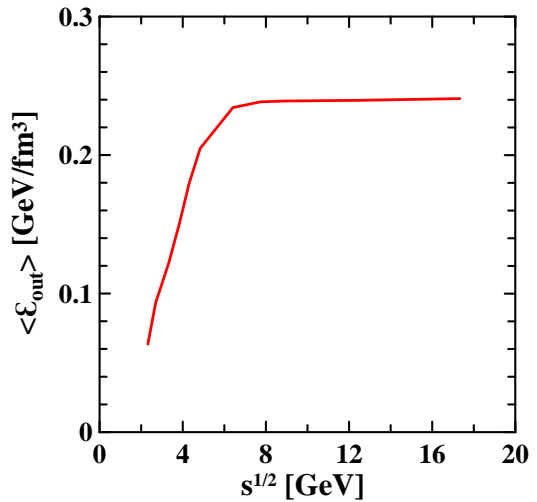}
\caption{(Color online)
Actual average freeze-out energy density
in central Pb+Pb collisions as a function of
invariant incident energy.
}
\label{fig1.1}
\end{figure}

In particular, this dynamical freeze-out results in a peculiar
incident-energy dependence of the actual freeze-out energy density
averaged over space--time evolution of the collision, 
$\langle\varepsilon_{\scr{out}}\rangle$, see Fig. \ref{fig1.1}. 
As seen,
$\langle\varepsilon_{\scr{out}}\rangle$ reveals fast rise at AGS energies and  saturation at the SPS energies.
This happens in spite of the fact that our freeze-out condition
involves only a single constant parameter%
---the ``triger'' freeze-out energy density 
$\varepsilon_{\scr{frz}}= 0.4$ GeV/fm$^3$---which was taken the same
for all incident energies with the exception of low incident
energies, for which we used lower values: 
$\varepsilon_{\scr{frz}}(2A \mbox{ GeV}) = 0.3$ GeV/fm$^3$ and 
$\varepsilon_{\scr{frz}}(1A \mbox{ GeV}) = 0.2$ GeV/fm$^3$. 
In our previous paper \cite{3FD}
we have performed only a rough analysis of this kind%
\footnote{This is why in the main text of Ref. \cite{3FD} we mentioned the value
of approximately 0.2 GeV/fm$^3$ for $\varepsilon_{\scr{out}}$ and in
the appendix
explained how the freeze-out actually proceeded. 
In terms of   Ref. \cite{3FD} 
($\varepsilon_{\scr{frz[1]}}$ and 
$\varepsilon_{\scr{frz[1]}}^{\scr{code}}$)
our present quantities are 
$\varepsilon_{\scr{frz}}=\varepsilon_{\scr{frz[1]}}^{\scr{code}}$
and 
$\varepsilon_{\scr{out}}=\varepsilon_{\scr{frz[1]}}$.}.
To find out the actual
value of $\varepsilon_{\scr{out}}$, we analyzed results of
actual simulations \cite{3FDfrz}.

The "step-like" behavior of $\langle\varepsilon_{\scr{out}}\rangle$
is a consequence of the freeze-out dynamics, as it was demonstrated in Ref.
\cite{3FDfrz}. At low (AGS) incident energies, the energy density
achieved at the border with vacuum, $\varepsilon^s$, is lower than
$\varepsilon_{\scr{frz}}$. Therefore, the surface freeze-out starts
at lower energy densities. It further proceeds at lower densities up
to the global freeze-out because the freeze-out front moves not
faster than with the speed of sound, like any perturbation in the
hydrodynamics. Hence it cannot overcome the
supersonic barrier and reach dense regions inside the expanding system.
With the incident energy rise the energy density achieved at the
border with vacuum gradually reaches the value of
$\varepsilon_{\scr{frz}}$ and then even overshoot it. If the
overshoot happens, the system first expands without freeze-out. The
freeze-out starts only when $\varepsilon^s$ drops to the value of
$\varepsilon_{\scr{frz}}$. Then the surface freeze-out occurs really
at the value $\varepsilon^s \approx \varepsilon_{\scr{frz}}$ and
thus the actual freeze-out energy density saturates at the value
$\langle\varepsilon_{\scr{out}}\rangle \approx
\varepsilon_{\scr{frz}}/2$, i.e. at the half fall from $\varepsilon^s$ to zero. 
This freeze-out dynamics is quite stable
with respect to numerics \cite{3FDfrz}.

\begin{figure}[thb]
\includegraphics[width=7.4cm]{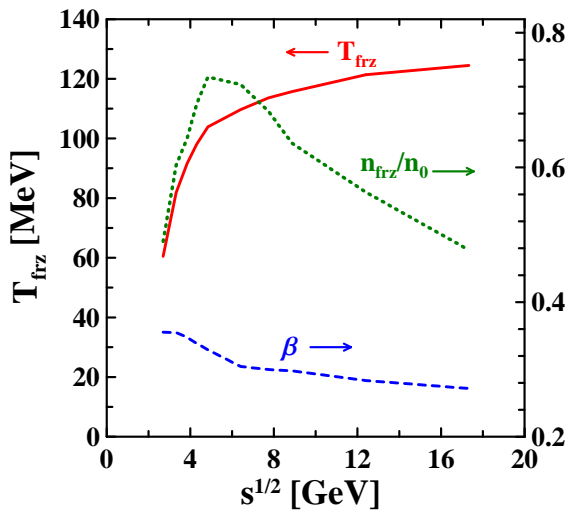}
\caption{(Color online)
Average temperature ($T_{\scr{frz}}$, see scale at the l.h.s.
axis), transverse velocity  ($\beta=v_{T}/c$, see scale at the r.h.s.
axis) and baryon density over the normal nuclear density
($n_{\scr{frz}}/n_0$, see scale at the r.h.s. axis) at the
freeze-out in central  Au+Au (at AGS energies) and Pb+Pb (at
SPS energies) collisions as a function of invariant
incident energy. Arrows point to the proper axis scale.
}
\label{fig1.2}
\end{figure}

Fig. \ref{fig1.2} clarifies this "step-like" behavior
in terms of the average
temperature, transverse velocity and baryon density achieved at the
freeze-out in central Au+Au (at AGS energies, impact parameter $b=2$ fm) 
and Pb+Pb (at
SPS energies, $b=2.5$ fm) collisions. The freeze-out temperature 
$T_{\scr{frz}}$ reveals a similar "step-like" behavior. 
At SPS energies the freeze-out temperatures in
Fig. \ref{fig1.2} are
noticeably lower than those deduced from hadron multiplicities in the
statistical model \cite{Andronic06,Cleymans06}.
The reason for this is as follows. Whereas
the statistical model assumes a single uniform fireball,
in the 3FD simulations at the late stage of the evolution the
system effectively consists of several ``fireballs'':
two overlapping fireballs (one baryon-rich and one baryon-free) 
at lower SPS energies and three fireballs 
(two baryon-rich and one baryon-free) at top SPS energies \cite{3FDfrz}. 
At the top SPS energies these
three fireballs turn out to be even spatially separated. Therefore,
whereas high multiplicities of mesons and antibaryons are achieved by
means of high temperatures in the statistical model, the 3FD model
explains them by an additional contribution of the baryon-free
fireball at a lower temperature. The freeze-out baryon density
$n_{\scr{frz}}$ exhibits a maximum at incident energies of
$E_{\scr{lab}}=$ 10$A$--30$A$ GeV
which are well within range of the planned FAIR in GSI.
This observation agrees with that deduced from the statistical model
\cite{Randrup06}, even  baryon density values in the maximum are
similar to those presented in Ref. \cite{Randrup06}.

\begin{figure}[thb]
\includegraphics[width=8.2cm]{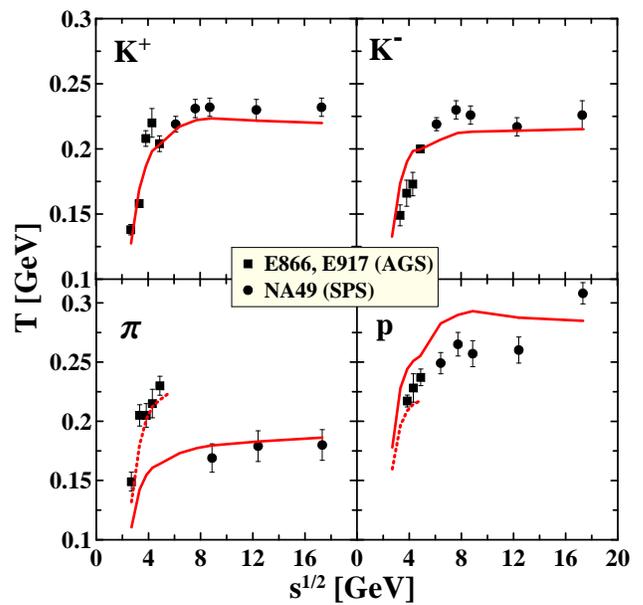}
\caption{(Color online)
Inverse-slope parameters of transverse-mass spectra of
kaons, pions and protons at midrapidity produced in central
Au+Au  and  Pb+Pb collisions as a function of
invariant incident energy.
Solid lines correspond to  purely exponential fit by Eq. (\ref{Ttr})
with $\lambda=0$, while dashed lines, as well as experimental data at AGS energies, correspond to fits with $\lambda=-1$
for pions and with $\lambda=1$ for protons.
Experimental data are from Refs. \cite{E866,mg:04,NA49,E917p,NA49p}.
}
\label{fig5.1}
\end{figure}

This dynamics of the freeze-out is of key importance for explanation
of the transverse-mass spectra. In particular, 
the "step-like" behavior of inverse slopes (effective temperatures) in
Fig. \ref{fig5.1} is related to 
the similar "step-like" behavior of $\langle\varepsilon_{\scr{out}}\rangle$
(see Fig. \ref{fig1.1}) or freeze-out temperature $T_{\scr{frz}}$
(Fig. \ref{fig1.2}).  
Of course, this relation is not direct, since the collective transverse
velocity also contributes to effective temperatures. However, since 
the collective transverse velocity at the freeze-out does not strongly change 
in the considered energy range (see Fig. \ref{fig1.2}), effective temperatures 
closely follow the shape 
of the $\langle\varepsilon_{\scr{out}}\rangle$ energy dependence. 
These inverse slopes $T$
were deduced by fitting the calculated spectra by the formula
\begin{eqnarray}
\label{Ttr}
\frac{d^2 N}{m_T \; d m_T \; d y} \propto
\left( m_T\right)^\lambda
\exp \left(-\frac{m_T}{T}  \right),
\end{eqnarray}
where $m_T$ and $y$ are the transverse mass and rapidity,
respectively, and $\lambda$ is a parameter which is taken different 
in different experimental fits, see, e.g., \cite{E866,E917p}.


Numerical problems, discussed in Ref. \cite{3FD}, prevented us from
simulations at RHIC energies. Already for the central Pb+Pb collision 
at the top SPS energy the code requires 7.5 GB of (RAM) memory.
At the top
RHIC energy, required memory is three order of magnitude higher, which is
unavailable in modern computers.

\section{Protons}
\label{Protons}
\begin{figure*}[thb]
\includegraphics[width=13.9cm]{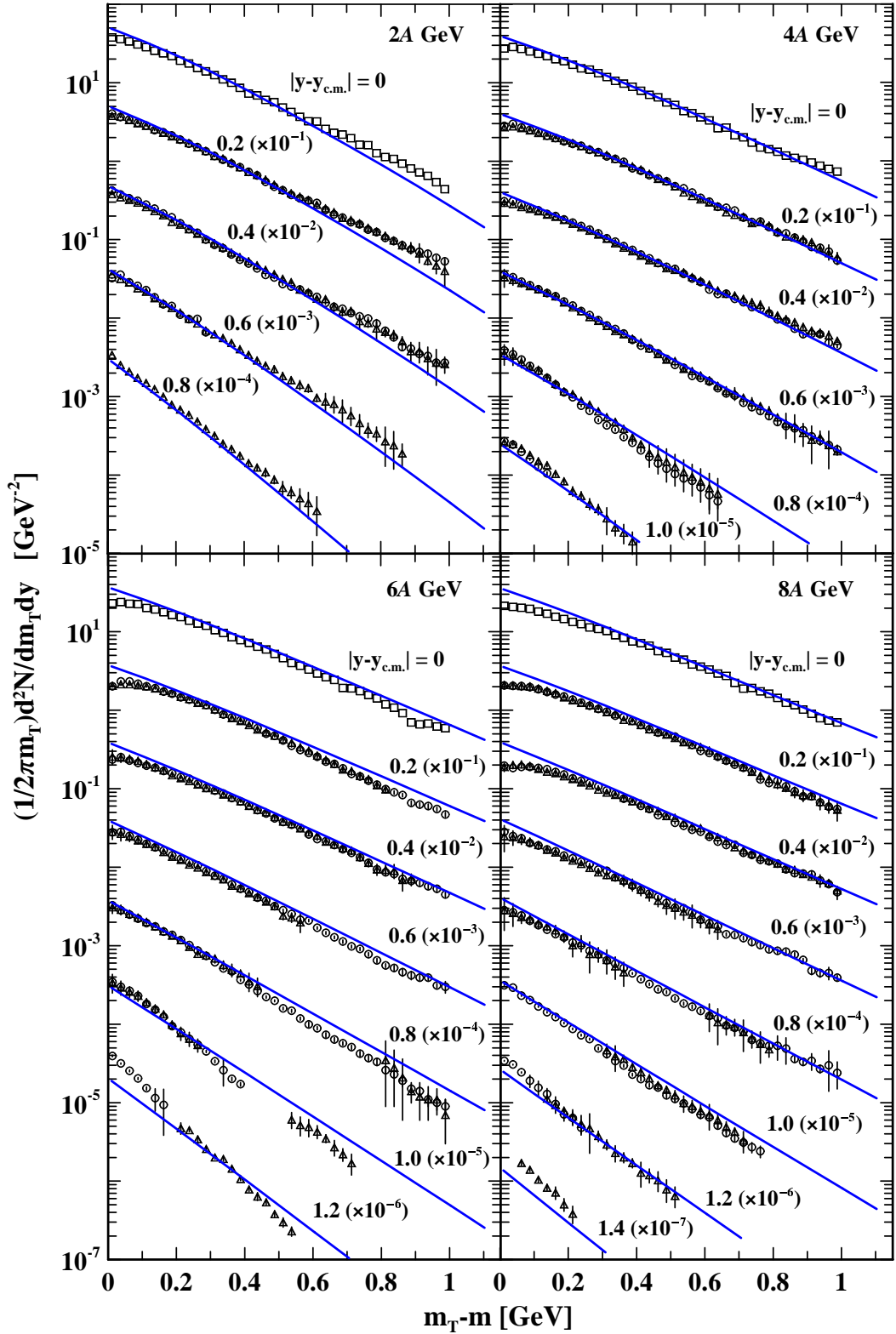}
\caption{(Color online)
Transverse-mass spectra of protons in various 
rapidity bins (centered at $|y-y_{c.m.}|$) from central Au+Au collisions at 
incident  
energies  $E_{\scr{lab}} =$ 2$A$, 4$A$, 6$A$, and 8$A$ GeV.
3FD results are presented for impact parameter $b=$ 2 fm. 
Midrapidity spectra  are shown unscaled,  while
every next data set and the corresponding curve (from top to bottom) are 
multiplied by additional 
factor 0.1. Data are from E895 Collaboration \cite{E895-2002}: 
open squares for midrapidity, circles for negative $(y-y_{c.m.})$
and triangles for positive  $(y-y_{c.m.})$. 
}
\label{fig2.1}
\end{figure*}
\begin{figure*}[thb]
\includegraphics[width=16.5cm]{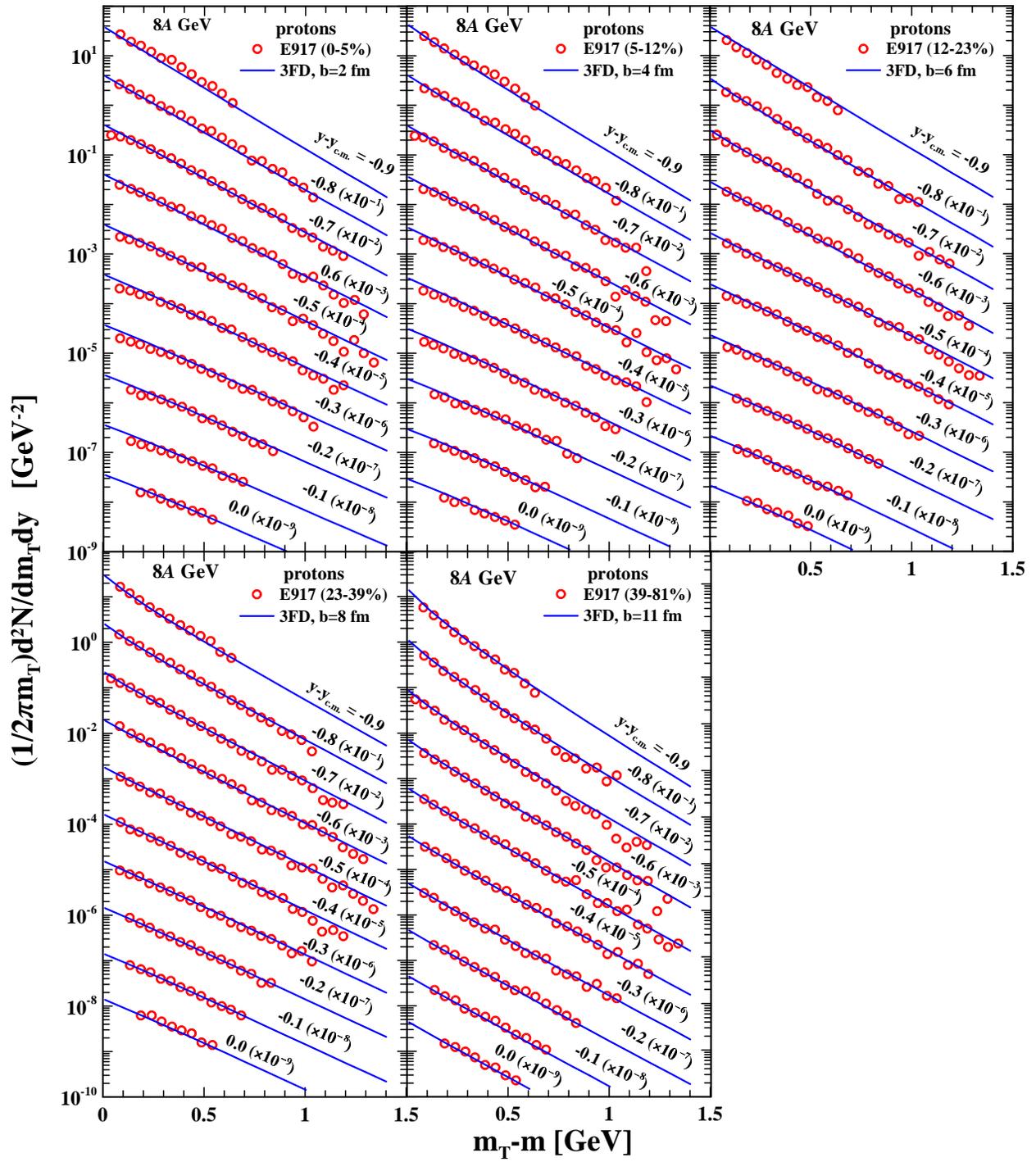}
\caption{ (Color online)
Transverse mass distributions of protons 
from Au(8$A$ GeV)+Au  collisions at various impact parameters ($b$)
and in various  rapidity bins (centered at $y-y_{c.m.}$). 
The most backward rapidity spectra are shown unscaled, while 
every next data set and the corresponding curve (from top to bottom) are 
multiplied by the additional factor 0.1. 
The percentage shows the fraction of the total reaction cross section, 
corresponding to experimental selection of events.
Experimental data are from  E917 Collaboration \cite{E917}. 
}
\label{fig2.2}
\end{figure*}
\begin{figure*}[thb]
\includegraphics[width=13.9cm]{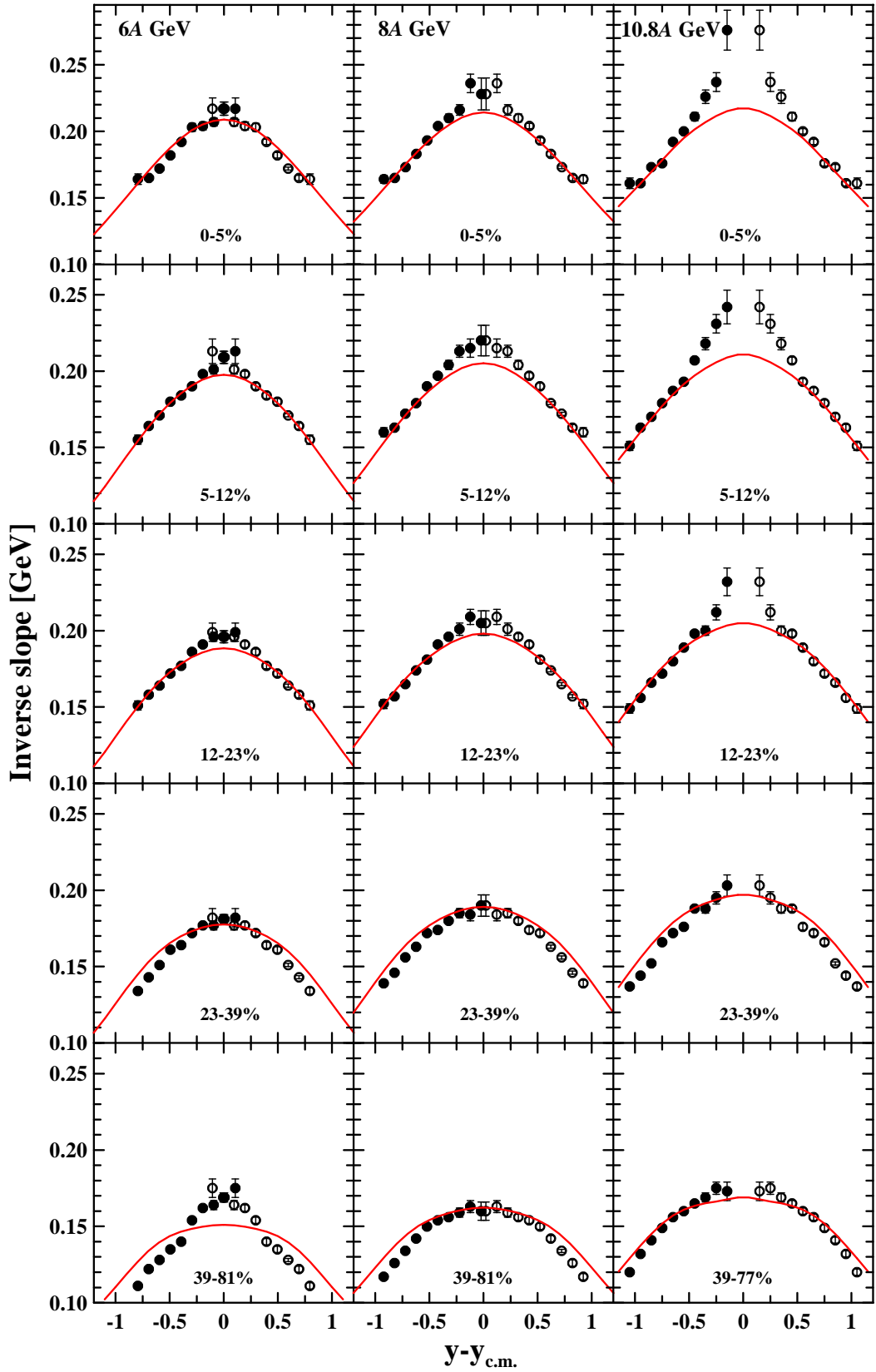}
\caption{(Color online)
Inverse-slope parameters of transverse-mass spectra of
protons produced in Au+Au  collisions at incident  
energies  $E_{\scr{lab}} =$  6$A$,  8$A$, 10.8$A$ GeV at various
centralities as a function of rapidity $y-y_{c.m.}$.
3FD results are presented for 
impact parameters $b=$ 2, 4, 6, 8, and 11 fm 
(from top row of panels to bottom one).  
The percentage indicates the centrality, i.e. the fraction of the
total reaction cross section,  
corresponding to experimental selection of events.
Experimental data are from E917 Collaboration \cite{E917p}. 
}
\label{fig2.2.1}
\end{figure*}
\begin{figure}[thb]
\includegraphics[width=7.9cm]{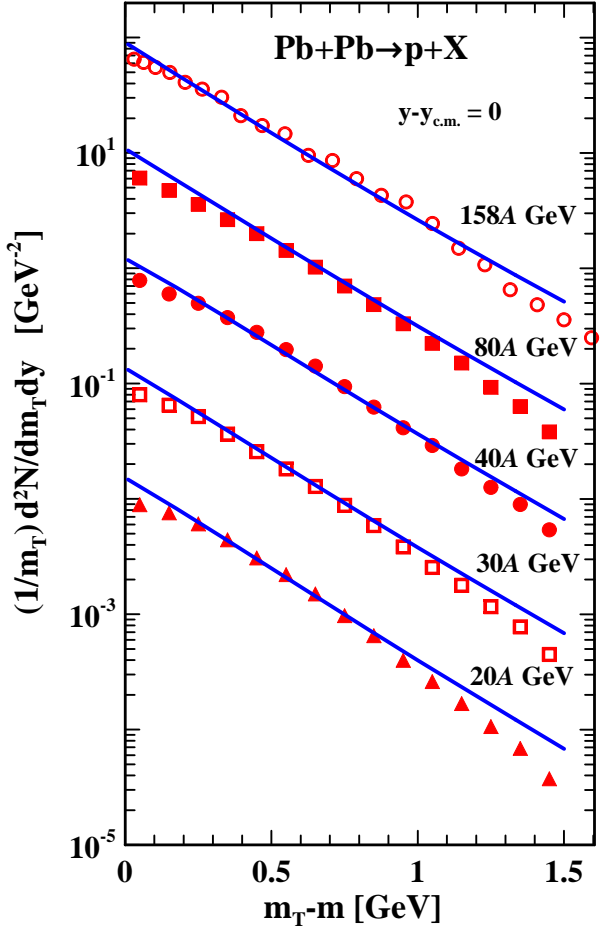}
\caption{(Color online)
Transverse-mass distributions of protons at the midrapidity
from central Pb+Pb collisions at incident energies 
$E_{\scr{lab}}=$ 158$A$, 80$A$, 40$A$ 30$A$ and 20$A$ GeV. 
3FD results are presented for impact parameter $b=$ 2.5 fm. 
The spectrum for 158$A$ GeV energy is shown unscaled, while 
every next data set and the corresponding curve (from top to bottom) are 
multiplied by the additional factor 0.1. 
NA49 experimental data are taken
from \cite{NA49-1} (158$A$ GeV), \cite{NA49-03} (80$A$ and 40$A$ GeV) and
\cite{mg:04} (30$A$ and 20$A$ GeV). 
}
\label{fig2.3}
\end{figure}
\begin{figure}[thb]
\includegraphics[width=7.9cm]{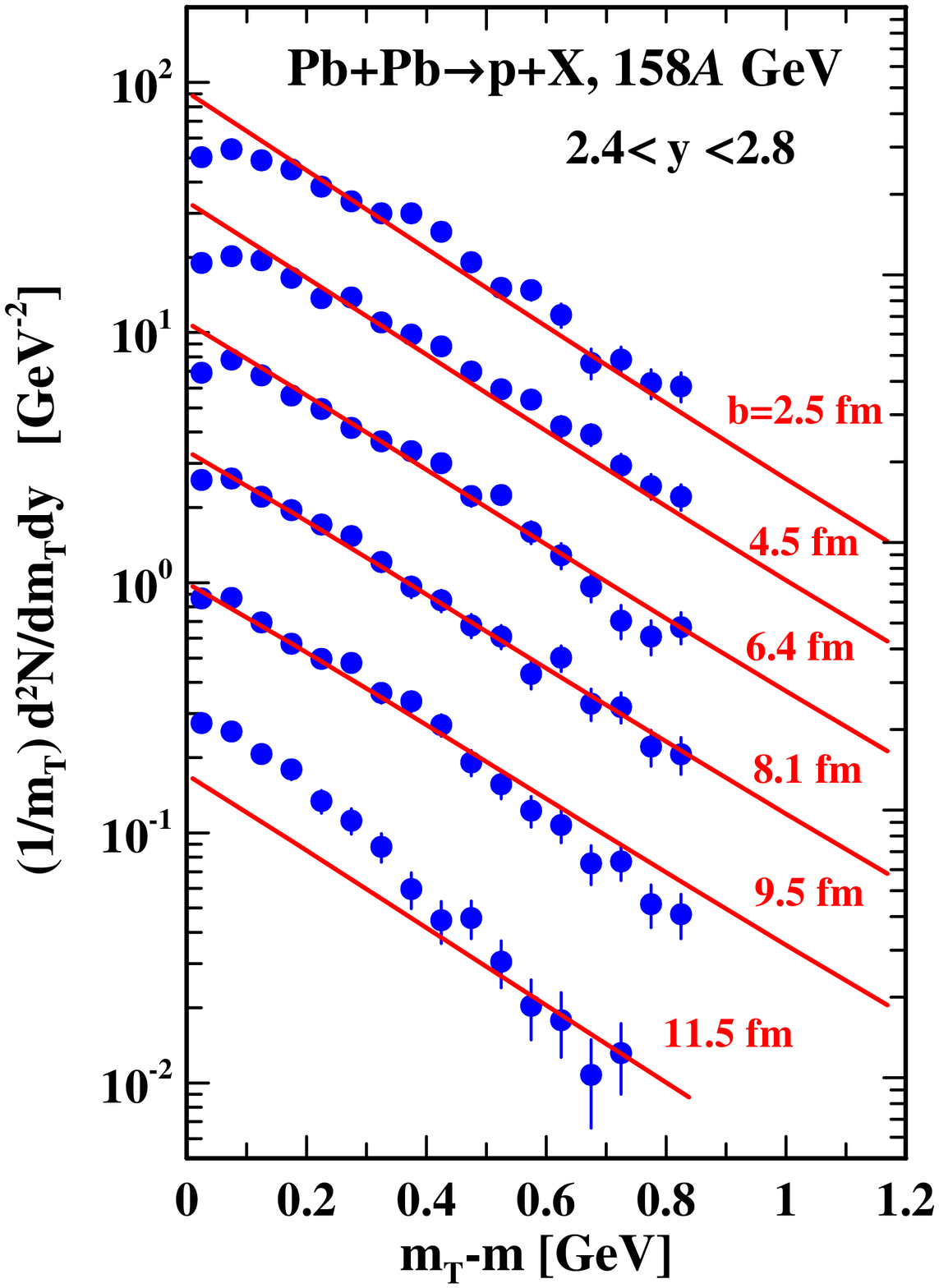}
\caption{(Color online)
Transverse-mass distributions of protons in the rapidity range 
$2.4<y<2.8$ from  Pb+Pb collisions at  incident energy 
$E_{\scr{lab}}=$ 158$A$  GeV and various centralities, i.e., impact
parameters $b$. 
The spectrum for central collisions is shown unscaled, while 
every next data set and the corresponding curve (from top to bottom) are 
multiplied by the additional factor 0.5. 
NA49 experimental data are taken
from Ref. \cite{NA49p}. 
3FD results are presented for rapidity $y=2.6$ and 
impact parameters estimated in Ref. \cite{NA49p}. 
}
\label{fig2.4}
\end{figure}
%
 
 
 

The 3FD model does not distinguish isotopic species. In particular,
the proton and neutron are considered to be identical, i.e. the
nucleon. Therefore, a spectrum of  protons is just associated with the
nucleon spectrum multiplied by the factor $Z/A$ with $Z$ and $A$ being
the charge and the mass number of colliding nuclei, respectively (in
this paper we consider only identical colliding nuclei). This
approximation is quite good at low incident energies, when the number
of produced secondary particles (mostly pions) is well smaller than
$A$. The particle production tends to restore isotopic
symmetry. Hence, the $Z/A$ approximation becomes worse at high
incident energies.  

Reproduction of available transverse-mass spectra of protons is
presented in Figs.  
\ref{fig2.1}--\ref{fig2.4}. We consider here only collisions of heavy
nuclei, since they offer favorable conditions for application of
the hydrodynamics.   
With few exceptions, overall reproduction of these spectra is quite
good in terms of both normalization and slope.

The first exception is the spectrum at the lowest considered energy of 2$A$ GeV, 
see Fig.  \ref{fig2.1}. The calculated spectrum turns out to be steeper  
than the experimental one
in all rapidity bins. This spectrum is calculated with our default choice of the freeze-out energy density
$\varepsilon_{\scr{frz}}(2A \mbox{ GeV}) = 0.3$ GeV/fm$^3$
accepted in our previous papers \cite{3FDpt,3FD,3FDflow}. 
Variation of $\varepsilon_{\scr{frz}}$ in the range from 0.1 to 0.6
GeV/fm$^3$ does not noticeably affects the slope of the spectrum. This
fact agrees with the above discussed dynamics of the freeze-out at low
energies. Indeed, the surface freeze-out occurs at lower energy
density which is the same independently of
$\varepsilon_{\scr{frz}}$. Only global freeze-out of the system
residue in the very end of the system disintegration produces slight
sensitivity to $\varepsilon_{\scr{frz}}$. This problem at the energy
of 2$A$ GeV indicates that the accepted model of the freeze-out is
poorly applicable at low incident energies.

\begin{figure*}[thb]
\includegraphics[width=12.9cm]{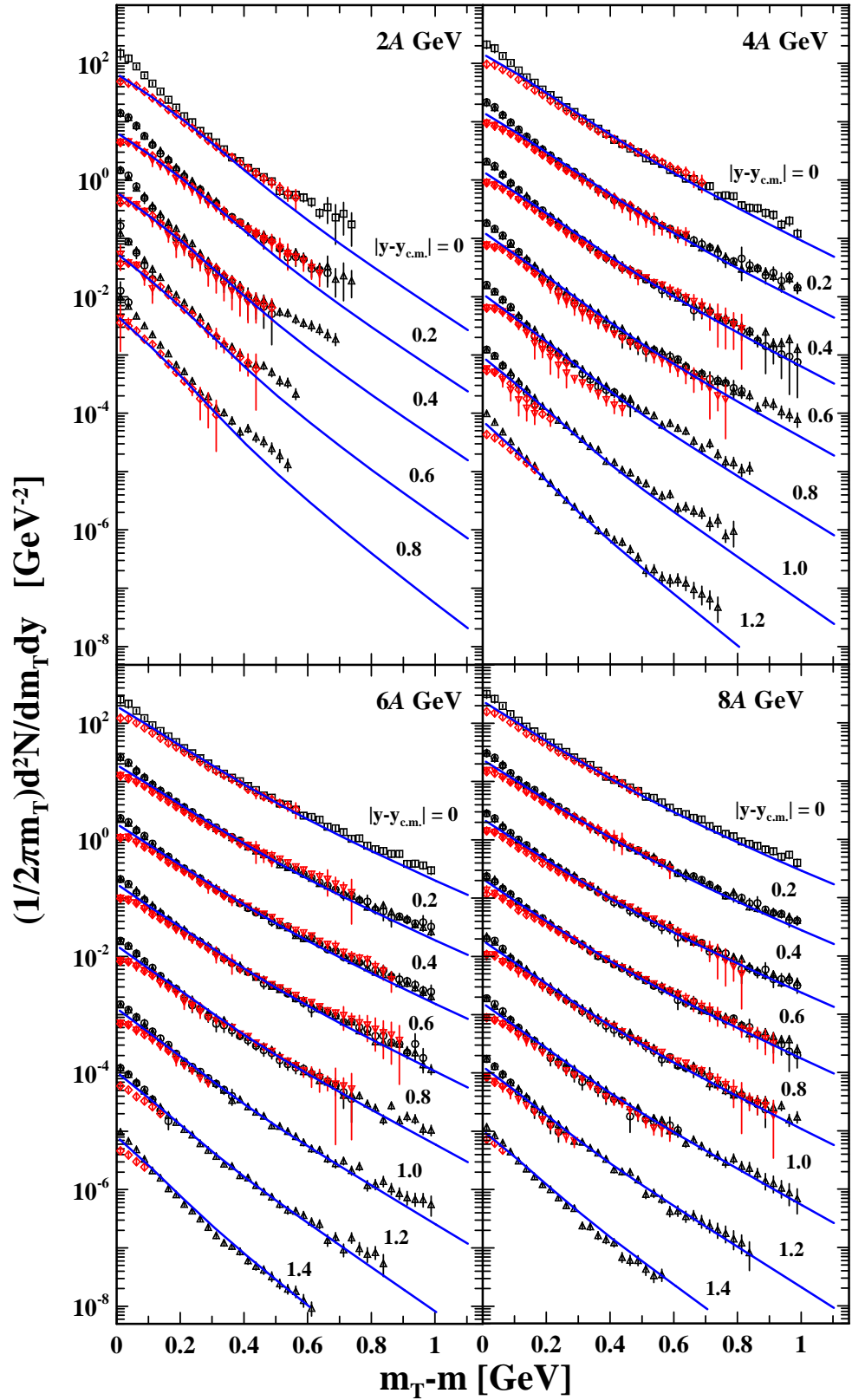}
\caption{(Color online)
Transverse-mass spectra of positive and negative pions in various
rapidity bins (centered at $|y-y_{c.m.}|$) for central Au+Au collisions at
incident  
energies  $E_{\scr{lab}} =$ 2$A$, 4$A$, 6$A$, and 8$A$ GeV.
3FD results (for $(\pi^+ + \pi^0 + \pi^-)/3$) are presented for impact
parameter $b=$ 2 fm.  
Midrapidity spectra  are shown unscaled,  while
every next data set and the corresponding curve (from top to bottom) are 
multiplied by the additional 
factor 0.1. Data are
from the E895 Collaboration  \cite{E895-2003}: 
squares for midrapidity $\pi^-$, 
triangles for positive-rapidity $\pi^-$, 
circles for negative-rapidity $\pi^-$, 
diamonds for midrapidity and positive-rapidity $\pi^+$,
inverted triangles for negative-rapidity $\pi^+$. 
}
\label{fig3.1}
\end{figure*}
\begin{figure}[thb]
\includegraphics[width=7.9cm]{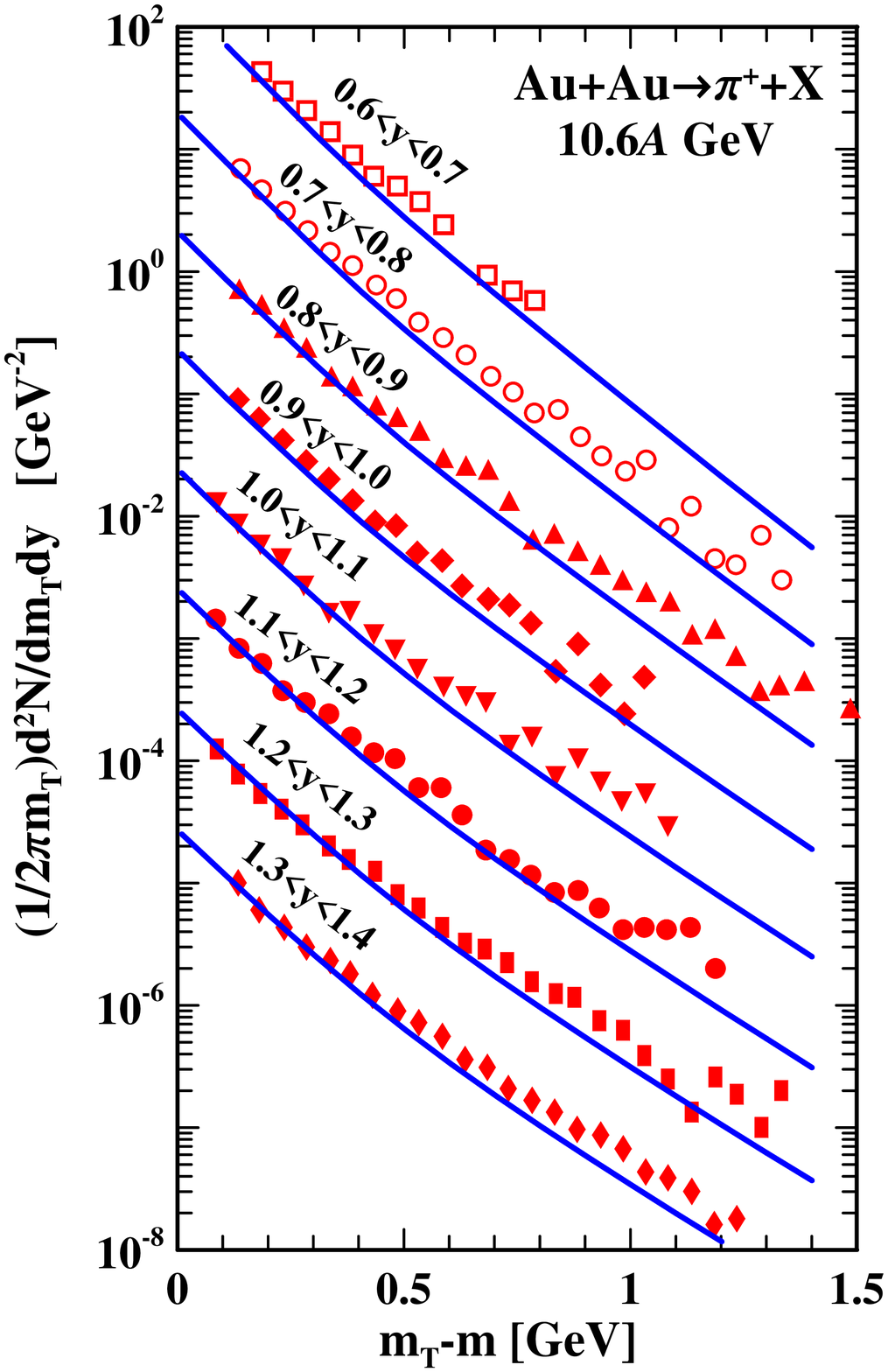}
\caption{(Color online)
Transverse-mass spectra of positive  pions in various 
rapidity bins from central Au+Au collisions at
incident  
energy  $E_{\scr{lab}} =$  10.6$A$ GeV.
3FD results (for $(\pi^+ + \pi^0 + \pi^-)/3$) are presented for impact
parameter $b=$ 2 fm.  
The most backward rapidity spectrum is shown unscaled, while 
every next data set and the corresponding curve (from top to bottom) are 
multiplied by the additional 
factor 0.1. Data are from the E802 Collaboration \cite{E802-1999}. 
}
\label{fig3.2}
\end{figure}

The 3FD 
model does not exhibit deviation from the exponential fall-off at
$(m_T-m)\lsim 0.2$ GeV, observed in the experiment, as it is most clearly seen in 
Figs. \ref{fig2.1} and \ref{fig2.4}. 
The same feature of the hydrodynamic calculation was earlier reported
in Ref. \cite{TLS01}. As it was shown \cite{TLS01}, a
post-hydro kinetic evolution (afterburner) is required to produce the observable two-slope
structure of the $m_T$-spectra. In our model such a post-hydro
evolution is absent. It is worthwhile to note that the above problem
is not an inevitable feature of any hydrodynamic calculation. For
instance, the deviation from the exponential fall-off was
reproduced in calculations by Kolb {\em et al.} \cite{Kol99}.

Traditionally, this deviation from the exponential fall-off at low
$(m_T-m)$ is associated with collective transverse expansion of the
system \cite{TLS01,Siemens}.  
In Ref. \cite{TLS01} it is described how the post-hydro cascade modify
this radial flow. Apparently the final-stage (post-hydro) Coulomb
interaction also distinctly affects the flow, accelerating or
decelerating it depending on the electric charge of the species.  It
is clear from the difference between spectra of positive and negative
pions, see Fig. \ref{fig3.1}. As seen from Fig. \ref{fig3.1}, negative
pions are  decelerated (their low-$m_T$ spectrum is enhanced) while
positive pions are accelerated (their low-$m_T$ spectrum is
suppressed). The same mechanism should also contribute to suppression
of low-$m_T$ spectrum of protons as compared to pure hydrodynamic
calculation. With rapidity moving off the midrapidity,  
$m_T$-spectra of protons reveal weaker suppression of their
low-$m_T$ parts, see Fig.  \ref{fig2.1}. This fact complies with the
Coulomb mechanism.  
Indeed, faster particles (in the frame, where the fireball generating
the field is at rest) spend shorter time in the Coulomb field and,
hence, acquire smaller additional momentum.  

Thus, keeping in mind that reproduction of the data in the low-$m_T$
region can be cured by application of a post-hydro dynamics, we see
that,  
in general, the 3FD model reasonably reproduces proton $m_T$-spectra
beyond extreme kinematic regions---rapidities being to far from the 
midrapidity  and too high $m_T$ (see Fig. \ref{fig2.1})---in not too
peripheral collisions of heavy nuclei (see Fig. \ref{fig2.4}). In the
above marginal regions applicability of hydrodynamics is
questionable. However, even in very peripheral collisions agreement
with data can be unexpectedly good, see Fig. \ref{fig2.2}.

Inverse-slope parameters [cf. Eq. (\ref{Ttr}) with $\lambda=1$], 
displayed in Fig. \ref{fig2.2.1}, summarize our results at AGS energies. 
The most pronounced disagreement with experimental temperatures occurs
in midrapidity region, especially at 10.8$A$ GeV incident energy. This
problem is directly related to the above-discussed poor reproduction
of low-$m_T$ suppression of the spectra.  
Apart from this problem, effective temperatures are well reproduced.

\section{Pions}
\label{Pions}

\begin{figure}[thb]
\includegraphics[width=7.9cm]{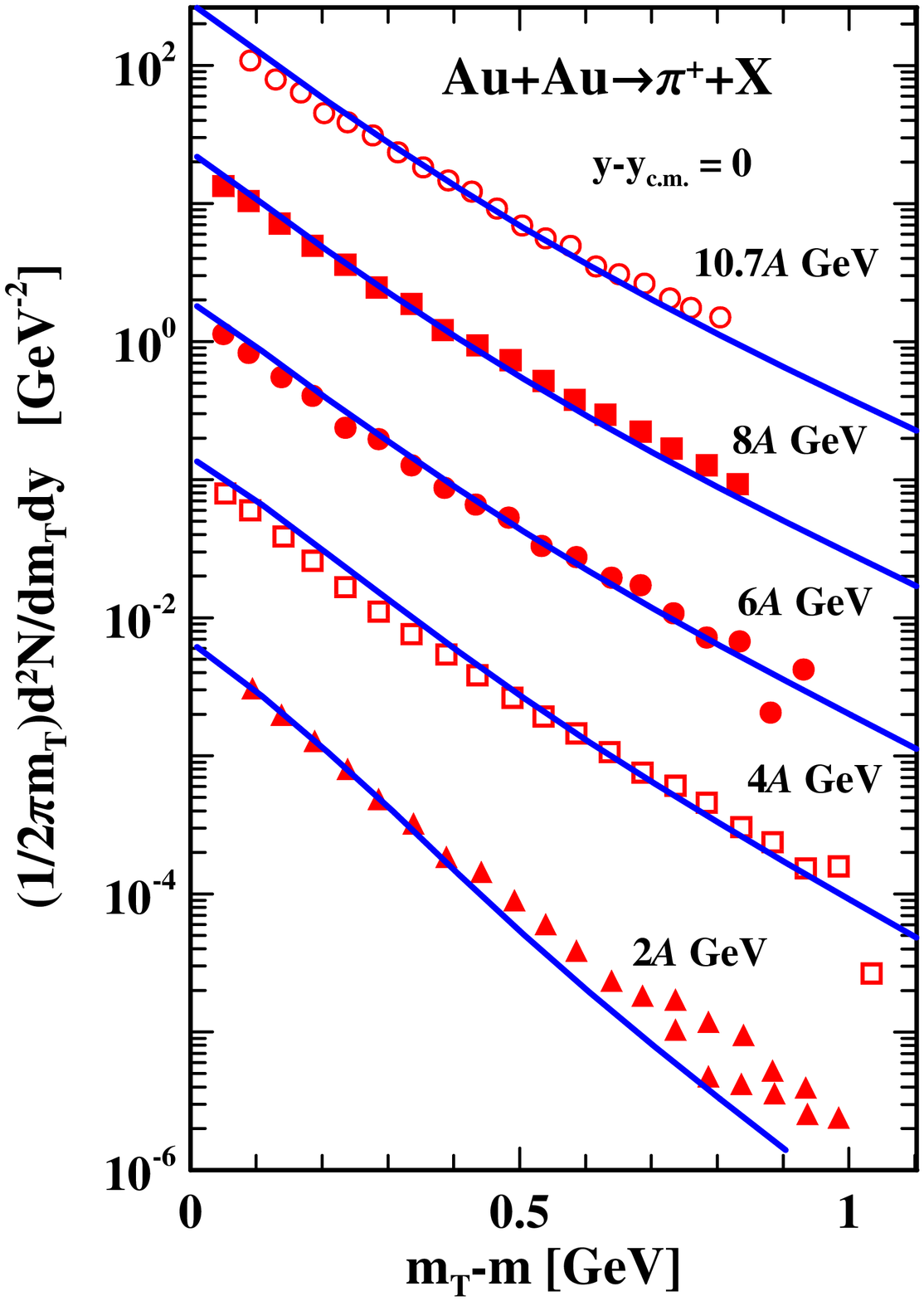}
\caption{(Color online)
Transverse-mass spectra of positive pions at midrapidity 
in central Au+Au collisions at incident  
energies  $E_{\scr{lab}} =$ 2$A$, 4$A$, 6$A$, 8$A$, and 10.7$A$ GeV.
3FD results (for $(\pi^+ + \pi^0 + \pi^-)/3$) 
are presented for impact parameter $b=$ 2 fm. 
Spectrum for 10.7$A$ GeV is shown unscaled, while  
every next data set and the corresponding curve (from top to bottom) are 
multiplied by the additional 
factor 0.1. Data are from the E866 Collaboration \cite{E866}. 
}
\label{fig3.3}
\end{figure}
\begin{figure}[thb]
\includegraphics[width=7.9cm]{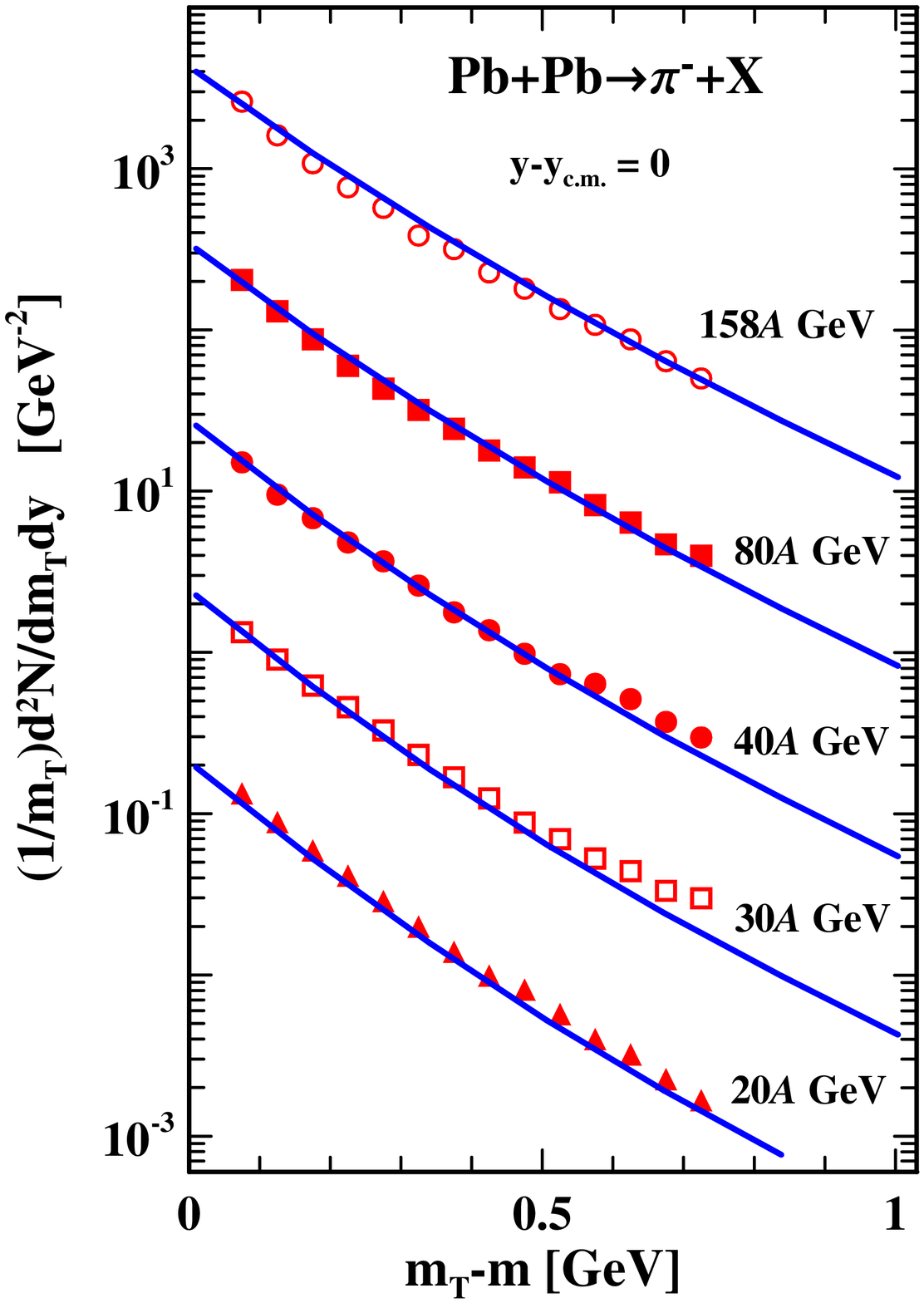}
\caption{(Color online)
Transverse-mass spectra of negative  pions at midrapidity in central
Pb+Pb collisions at various incident  
energies  $E_{\scr{lab}} =$  20$A$, 30$A$, 40$A$, 80$A$, and 158$A$  GeV.
3FD results (for $(\pi^+ + \pi^0 + \pi^-)/3$) are presented for impact
parameter $b=$ 2.5 fm.  
The spectrum for 158$A$ GeV energy is shown unscaled, while 
every next data set and the corresponding curve (from top to bottom) are 
multiplied by the additional factor 0.1. 
Data are from  the NA49 Collaboration  \cite{NA49}. 
}
\label{fig3.4}
\end{figure}

Comparison of calculated transverse-mass spectra of pions with available data 
is presented in Figs. \ref{fig3.1}--\ref{fig3.4}. 
Since the 3FD model does not distinguish isotopic species, we assume
that numbers (and spectra) of $\pi^+$, $\pi^-$ and $\pi^0$ are
identical and equal $1/3$ of the total number (spectrum) of pions. In
view of this approximation, if calculated spectra are situated between measured
spectra of positive and negative pions, that would mean 
a good agreement of our calculations with data.  
Contrary to protons, this approximation becomes better applicable at
high incident energies, when abundant production of secondary
particles results in partial restoration of isotopic symmetry of the
system.  
As it is illustrated in Fig. \ref{fig3.1}, 
spectra of positive and negative
pions indeed become more and more similar with the incident energy
rise.

Similarly to the proton case, agreement with data is the worst at the
lowest considered incident energy of 2$A$ GeV, see
Fig. \ref{fig3.1}. There our calculated spectra closely follow
experimental spectra for positive pions, while they should lie in
between  $\pi^+$ and $\pi^-$ spectra. Above this energy,  
overall reproduction of pion spectra is quite good both in
normalization and slope.  
It is even better than that for protons. 
The calculated pion  spectra also reproduce the rapidity
(Figs. \ref{fig3.1} and \ref{fig3.2}) and incident-energy
(Figs. \ref{fig3.3} and \ref{fig3.4}) dependence of the data.  
Note that this agreement is achieved with the unified chemical-thermal 
freeze-out. Of course, two different freeze-out criteria for 
chemical and thermal processes could slightly improve this agreement 
with the experiment. However, that would be too high price for so 
tiny improvement.

\section{Kaons}
\label{Kaons}
\begin{figure}[thb]
\includegraphics[width=7.9cm]{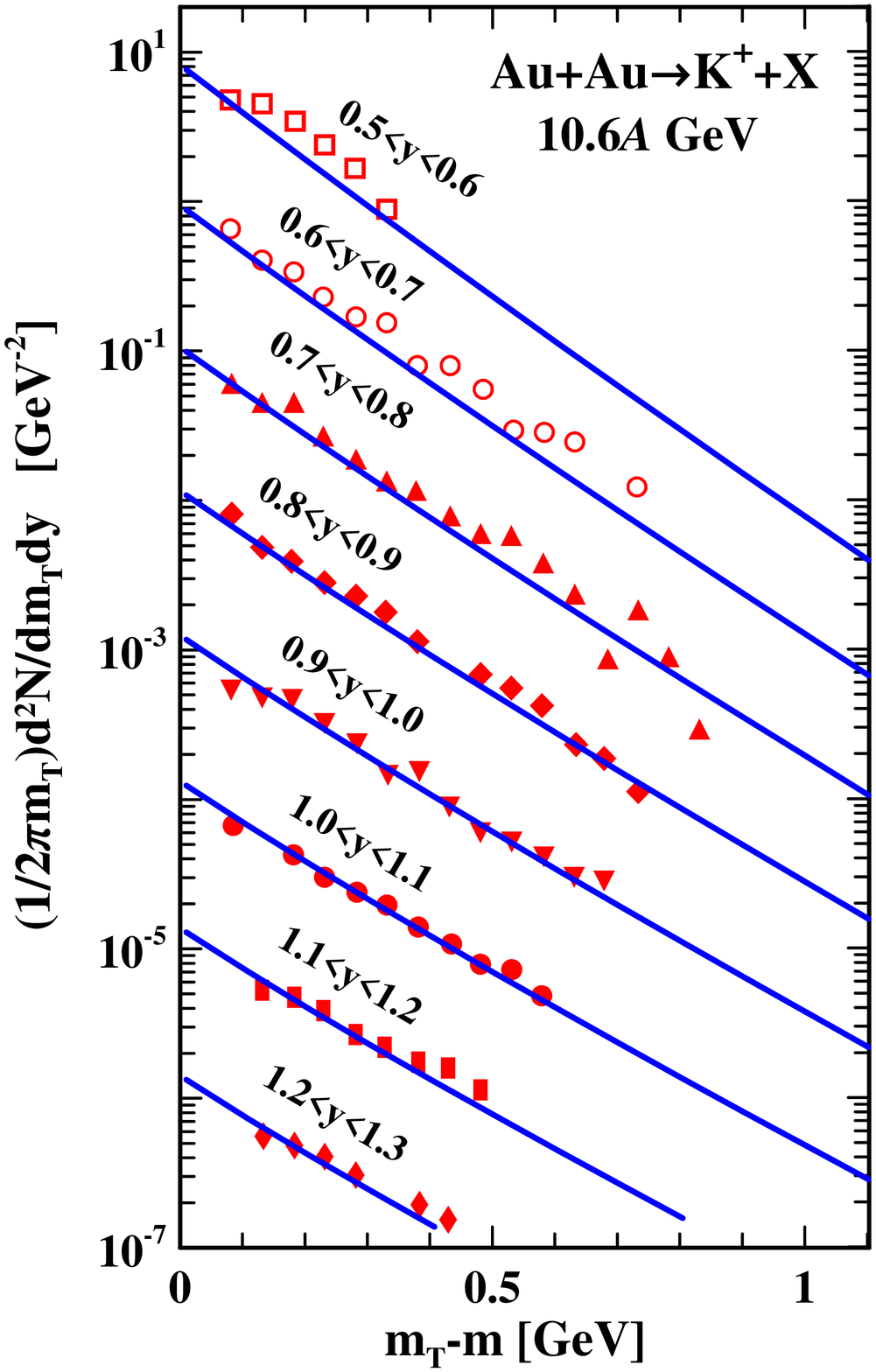}
\caption{(Color online)
Transverse-mass spectra of positive  kaons in various
rapidity bins in central Au+Au collisions at
incident  
energy  $E_{\scr{lab}} =$  10.6$A$ GeV.
3FD results  are presented for impact
parameter $b=$ 2 fm.  
The most backward rapidity spectrum is shown unscaled, while 
every next data set and the corresponding curve (from top to bottom) are 
multiplied by the additional 
factor 0.1. Data are from  the E802 Collaboration  \cite{E802-1999}.
}
\label{fig4.1}
\end{figure}
\begin{figure}[thb]
\includegraphics[width=7.9cm]{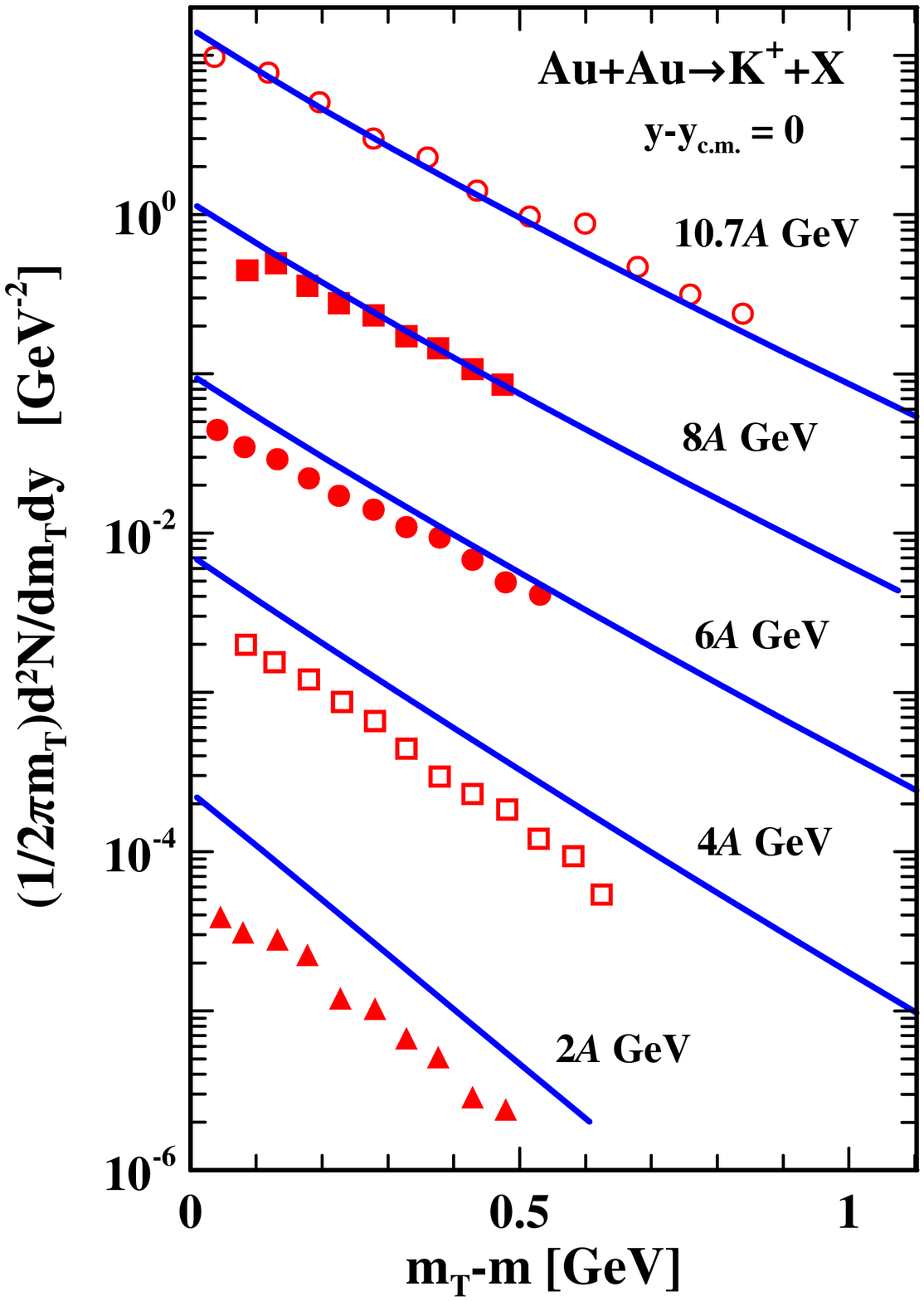}
\caption{(Color online)
Transverse-mass spectra of positive kaons at midrapidity 
in central Au+Au collisions at incident  
energies  $E_{\scr{lab}} =$ 2$A$, 4$A$, 6$A$, 8$A$, and 10.7$A$ GeV.
3FD results are presented for impact parameter $b=$ 2 fm. 
Spectrum for 10.7$A$ GeV is presented unscaled, while  
every next data set and the corresponding curve (from top to bottom) are 
multiplied by the additional 
factor 0.1. Data are from  the E866 Collaboration  \cite{E866}. 
}
\label{fig4.2}
\end{figure}
\begin{figure}[thb]
\includegraphics[width=7.9cm]{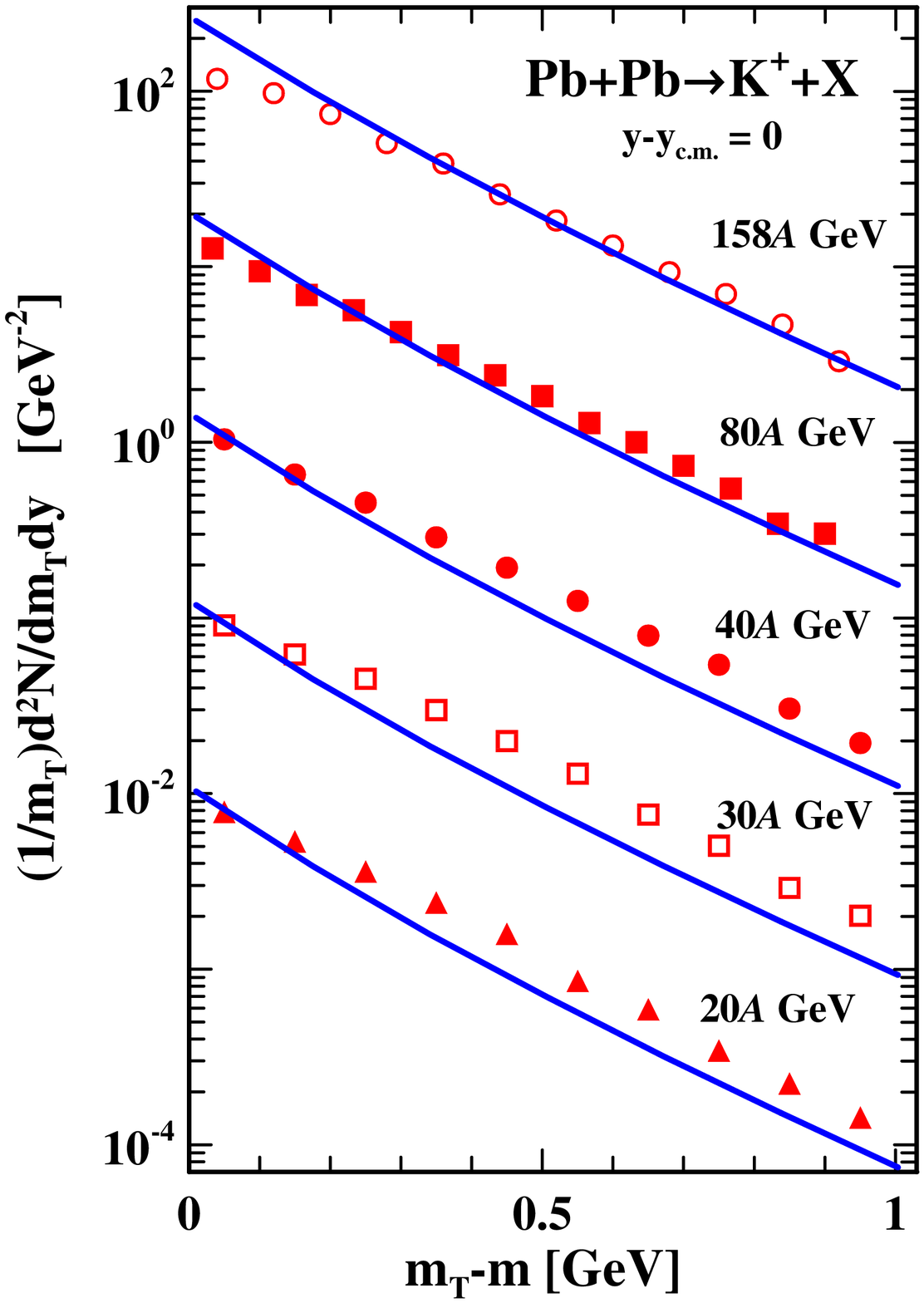}
\caption{(Color online)
Transverse-mass spectra of positive  kaons at midrapidity in central
Pb+Pb collisions at various incident  
energies  $E_{\scr{lab}} =$  20$A$, 30$A$, 40$A$, 80$A$, and 158$A$  GeV.
3FD results are presented for impact
parameter $b=$ 2.5 fm.  
The spectrum for 158$A$ GeV energy is shown unscaled, while 
every next data set and the corresponding curve (from top to bottom) are 
multiplied by the additional factor 0.1. 
Data are from  the NA49 Collaboration  \cite{NA49}. 
}
\label{fig4.3}
\end{figure}

\begin{figure}[thb]
\includegraphics[width=7.9cm]{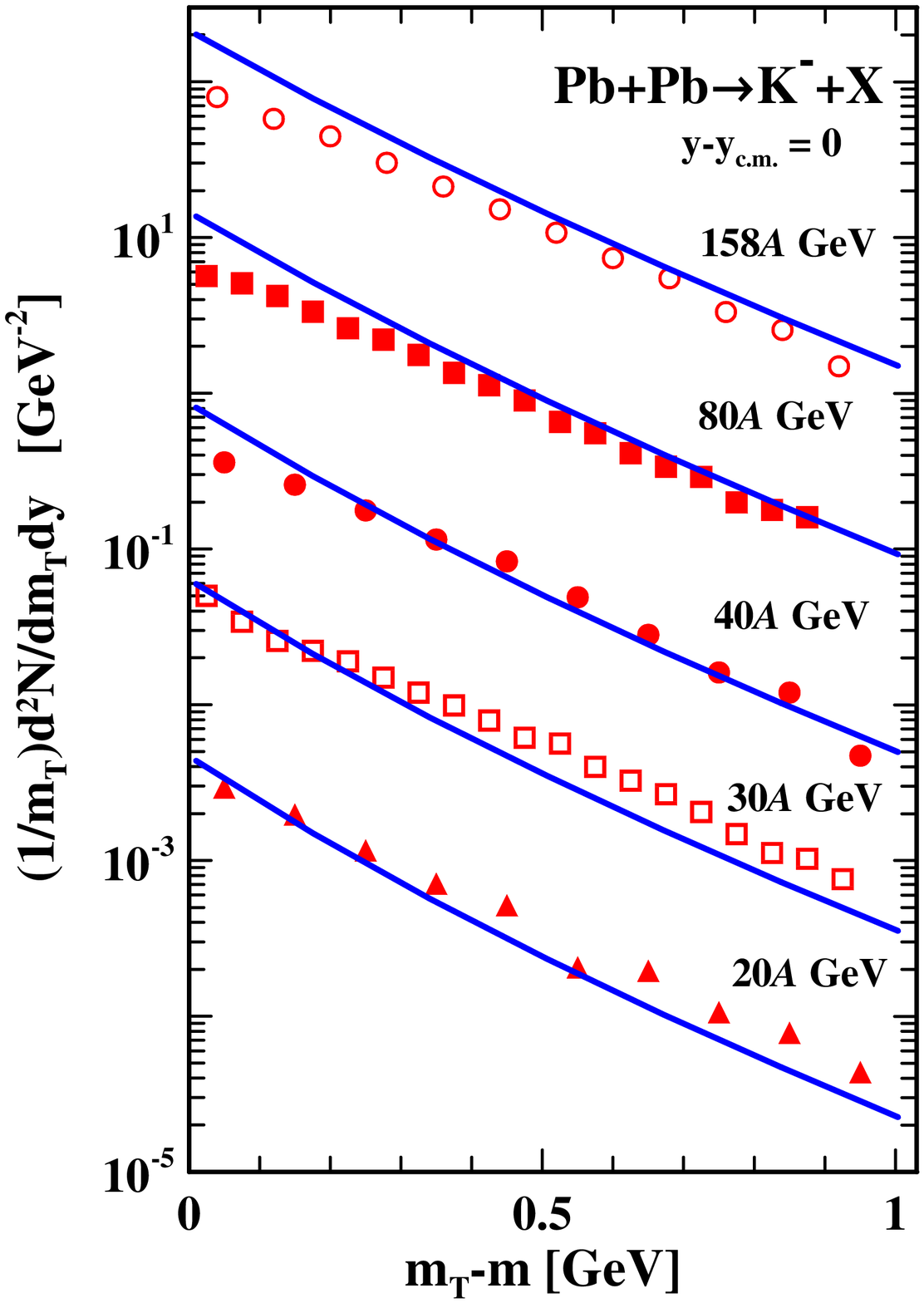}
\caption{(Color online)
The same as in Fig. \ref{fig4.3} but for negative  kaons. 
}
\label{fig4.4}
\end{figure}

Upon demonstrating that calculated proton and pion transverse-mass
spectra are in reasonable agreement with available data we proceed to
kaon spectra.  
The "step-like" incident energy dependence of the the inverse-slope
parameter ($T$, cf. Eq. (\ref{Ttr})) of these spectra was interpreted
as a signal of  
the deconfinement phase transition \cite{Gorenstein03,Mohanty03}. Moreover, 
microscopic transport
models, based on hadronic degrees of freedom, failed to reproduce
the observed behavior of the kaon inverse slope \cite{Bratkovskaya,Wagner}.

Comparison of calculated of transverse-mass spectra of kaons with 
available data in collisions of heavy nuclei is presented in Figs. 
\ref{fig4.1}--\ref{fig4.4}. Overestimation of kaon normalization at
low incident energies  ($E_{\scr{lab}} \lsim 6A$ GeV) is not
surprising, see Fig.  \ref{fig4.2}. 
This is not a result of the unified chemical-thermal freeze-out. 
The reason is that 
we use the EoS based on grand canonical ensemble, therefore
production of rare particles should be overestimated at low energies.   
This normalization can be easily corrected by introducing a
suppression factor $\gamma_s$, taking into account associative
production of strangeness.  
Apart from this normalization,  
overall reproduction of kaon spectra is quite satisfactorily. 
Separate treatment of the chemical and thermal freeze-outs could 
certainly improve the normalization of kaon spectra at SPS energies. 
However, to do this 
we would have to introduce different freeze-out criteria for $K^+$ 
and $K^-$. This is distinctly seen, e.g., at the example of $K^+$ 
and $K^-$ spectra at $E_{\scr{lab}}=80A$ GeV, 
see Figs. \ref{fig4.3} and \ref{fig4.4}. The $K^+$ spectrum requires 
somewhat earlier freeze-out in order to increase the normalization.  
While the $K^-$ spectrum demands for later freeze-out to reduce 
the normalization. In fact, there are physical arguments in favor of 
this sequence of $K^+$ and $K^-$ freeze-outs. We prefer a cruder 
description of observables to introduction of additional fitting 
parameters.

Dependence on the
rapidity is reproduced,  see Fig.  
\ref{fig4.1}, as well as evolution of slopes with the incident energy
variation, see Figs. \ref{fig4.1}--\ref{fig4.4}.  
Fig. \ref{fig5.1} demonstrates that inverse slopes of kaon spectra are
indeed reasonably reproduced.  
This inverse slopes were deduced by fitting the calculated spectra by
formula (\ref{Ttr}) with $\lambda=0$ (purely exponential fit). 
Moreover, the pion and proton effective
temperatures also reveal saturation at SPS energies, if they are
deduced from the same purely exponential fit with $\lambda=0$.
Though the purely exponential fit with $\lambda=0$
does not always provide the best fit of the spectra, it allows a
systematic way of comparing spectra at different incident energies.
In order to comply with experimental
fits at AGS energies (and hence with displayed experimental points
displayed by squares), 
we also present results of fits with $\lambda=-1$ for pions
and with $\lambda=1$ for protons. 
The most pronounced disagreement with experimental temperatures takes
place for protons, while reproduction of the proton spectra themselves 
(see Fig. \ref{fig2.3}) does not look bad. 

\section{Summary}
\label{Summary}



Transverse-mass 
spectra of protons, pions and kaons in wide ranges of incident
energies (from AGS  
to SPS), rapidity bins and centralities have been analyzed within the
3FD model.  
These spectra were computed with the same set of model parameters 
as that summarized in Ref. \cite{3FD}. In particular, the hadronic EoS 
\cite{gasEOS} with incompressibility $K=210$ MeV was used. 
We considered here only collisions of heavy nuclei, since they offer
favorable conditions for application of the hydrodynamics.  
It was demonstrated that with few exceptions 
the 3FD model reasonably reproduces these spectra 
in the range of incident energies 
4$A$ GeV $\lsim E_{\scr{lab}} \lsim$ 160$A$ GeV. 
The exceptions concern 
extreme kinematic regions---rapidities being to far from midrapidity
and too high $m_T$---and too peripheral collisions of heavy nuclei. In
the above marginal regions applicability of the hydrodynamics is
questionable. However, even in very peripheral collisions agreement
with data can be unexpectedly good.  

The 3FD model reproduces inverse slopes of $m_T$-spectra, in particular, 
the "step-like" dependence of kaon inverse slopes on the incident energy. 
This hydrodynamic explanation of the transverse-mass 
spectra and "step-like"
behavior of effective temperatures implies that 
at considered incident energies 
a heavy nuclear 
system really reveals a hydrodynamic motion during its expansion.

In fact, microscopic transport and hydrodynamic  models do not
contradict each other. The fact that they all successfully describe
many observales in collisions heavy nuclei in spite of big differences
between these models  
suggests that detailed information on cross sections is unimportant
for these reactions. The collision frequency is already high enough in
order to make hydrodynamics to be applicable. This conjecture has
already been put forward in Ref. \cite{HS95}. In particular, transport
\cite{Bratkovskaya,Wagner} and hydrodynamic models 
well describe pion transverse-mass 
spectra of pions in wide range of incident energies. 
Therefore, the failure of microscopic transport
models  \cite{Bratkovskaya,Wagner} to reproduce
the observed behavior of the kaon inverse slopes may be interpreted as
a signature  
that kaon and antikaon interaction cross sections in microscopic
models are not big enough for kaons to be captured by matter in a
common flow.  
In Ref. \cite{Bratkovskaya} a kind of such cross-section enhancement
(Cronin effect) 
was tested against the experimental data. The resulting effect of that
enhancement was found insufficient to explain the observed
discrepancy. 
Another possibility is that multi-body collisions are important in the
transport. This was checked within the Giessen
Boltzmann-Uehling-Uhlenbeck (GiBUU) model \cite{Larionov07},
where three-body interactions were included in simulations.  
It was found that the three-body collisions indeed result in good
reproduction of all transverse-mass spectra \cite{Larionov07}.

Returning to the question if the considered "step-like"
behavior of effective temperatures is a signal 
of a phase transition into the QGP, we have to admit that this is 
not quite clear as yet. It depends on the nature of the freeze-out 
parameter $\varepsilon_{\scr{frz}}=0.4$ GeV/fm$^3$ which should be further 
clarified. 
The EoS is not of prime importance for this 
behavior. The only constraint on the EoS is that it should be
in some way reasonable. 
Moreover, our preliminary results indicate
that a completely different (from that used in this work) 
EoS \cite{Toneev06} with 1st order phase
transition to the QGP still reasonably reproduces this "step-like"
behavior even in spite of that it fails to describe a large
body of other data. This happens 
because the same freeze-out pattern is accepted there. 

This kind of the freeze-out has its implications for observation of
fluctuations. A number of observables were suggested, which are based on
various fluctuations: transverse-momentum fluctuations
\cite{Stephanov:1999zu}, electric-charge fluctuations
\cite{Jeon:2000wg,Asakawa:2000wh}, 
baryon-charge--strangeness correlations \cite{Koch:2005vg},
fluctuations of the $K/\pi$ ratio  \cite{Koch:2005pk}; for the recent
review see  Ref. \cite{Koch08}. These observables distinguish between
the hadronic and quark--gluon phases. However, available experimental
data on fluctuations: on $p_T$ fluctuations 
\cite{Sako:2004pw,Appelshauser:1999ft,Rybczynski:2008cv,Grebieszkow:2007xz}
and positive-to-negative charge-ratio fluctuations
\cite{Alt:2004ir,Appelshauser:2004ms}, at SPS energies well agree
with the hadronic predictions. Only the $K/\pi$ fluctuations
\cite{Alt:2008ca} reveal a nontrivial behaviour with the incident energy.

The fact that the freeze-out takes place at rather low values of
energy density and temperature (see Figs. \ref{fig1.1} and
\ref{fig1.2}) explains why we observe only hadronic fluctuations at
the freeze-out, even if the compressed matter was in the quark--gluon
phase. All the quark--gluon fluctuations get dissolved during the long
path of the system from hadronization to freeze-out. Fluctuations of
the $K/\pi$ ratio can be an exception from this rule, since this
signal is the most robust. A $K^+$ meson can disappear only if it
meets a particle of the opposite strangeness ($\Lambda$, $\Sigma$ or
$K^-$). If the strange system is relatively dilute at the hadronic
stage of the expansion, as it is the case at lower SPS energies, the
probability to meet a particle of the opposite strangeness is
relatively low. Therefore, the signal from the quark--gluon phase
partially survives. At higher energies (top SPS energies) the strange
system becomes already dense enough to destroy this signal. This
mechanism can explain a nontrivial energy dependence of the $K/\pi$
fluctuations, observed by the NA49 Collaboration \cite{Alt:2008ca}. 
It implies violation of the chemical equilibrium for
strange particles at the late stage of the expansion. 
Therefore, it is beyond the frame of the present version of
the 3FD model, where $K^+$ meson abundance immediately follow changes
of densities of the matter.  
If this explanation is true, fluctuations of the ratio 
$K^+/\pi^+=N_{K^+}/N_{\pi^+}$ should reveal even stronger energy
dependence than those of the ratio
$K/\pi=(N_{K^+}+N_{K^-})/(N_{\pi^+}+N_{\pi^-})$, where $N_a$ is the
observed yield of $a$-particles.  
The same mechanism can be behind the horn effect in the excitation
function of the $K^+/\pi^+$ ratio \cite{na49.K/pi}.

\section*{Acknowledgements}


We are grateful to I.N. Mishustin, L.M. Satarov,
V.V.~Skokov, V.D. Toneev,  and D.N. Voskresensky for fruitful
discussions.
This work was supported 
the Deutsche
Forschungsgemeinschaft (DFG project 436 RUS 113/558/0-3), the
Russian Foundation for Basic Research (RFBR grant 06-02-04001 NNIO\_a),
Russian Federal Agency for Science and Innovations
(grant NSh-3004.2008.2).

\end{document}